\documentclass[10pt, twocolumn, showkeys, showpacs, nofootinbib]{revtex4-1}
\usepackage[english]{babel}
\usepackage{amsmath,amssymb,amsfonts}
\usepackage{graphicx}
\usepackage{hyperref}

\begin{document}

\title{{On applicability of differential mixing rules\\ for
statistically homogeneous and isotropic dispersions}}

\author{A. K. Semenov}
\email{andrey.k.semenov@gmail.com}
\address{Department of Theoretical Physics and Astronomy,\\
Odessa I.I.Mechnikov National University,
 2 Dvoryanskaya St., Odessa 65026, Ukraine}

\begin{abstract}
The classical differential mixing rules are assumed to be independent effective-medium approaches, applicable to certain classes of systems. In the present work, the inconsistency of differential models for macroscopically homogeneous and isotropic systems is illustrated with a model for the effective permittivity of simple dielectric systems of impenetrable balls. The analysis is carried out in terms of the compact group approach reformulated in a way that allows one to analyze the role of different contributions to the permittivity distribution in the system. It is shown that the asymmetrical Bruggeman model (ABM) is physically inconsistent since the electromagnetic interaction between previously added constituents and those being added is replaced by the interaction of the latter with recursively formed effective medium. The overall changes in the effective permittivity due to addition of one constituent include the contributions from both constituents and depend on the system structure before the addition. Ignoring the contribution from one of the constituents, we obtain generalized versions of the original ABM mixing rules. They still remain applicable only in a certain concentration ranges, as is shown with the Hashin-Shtrikman bounds. The results obtained can be generalized to macroscopically homogeneous and isotropic systems with complex permittivities of constituents.
\end{abstract}



\maketitle

\section{Introduction}\label{sec:intro}
A big variety of analytical methods and approaches have been
developed to study  electrophysical characteristics of disperse
systems and mixtures~\cite{Banhegyi1986, dudney1989, Nan1993,
Mackay00, Sushko2013}. However, it is often unclear which one is
applicable to a given system~\cite{Banhegyi1986, Shutko1982,
Bordi02, Nelson2005}. Even if a particular approach can be used
for one type of systems, it can be unapplicable to other similar
systems. Of the most used, but arguable approaches are
differential models~\cite{Bruggeman1935a, Bruggeman1935b,
Hanai1960, Hanai1980, Sen1981, Hanai1986}.

The first differential mixing rule was developed by Bruggeman for
the effective permittivity of mixtures~\cite{Bruggeman1935a,
Bruggeman1935b}; it is now known as the asymmetrical Bruggeman
model (ABM). Later, the ABM was generalized by Hanai and other
authors~\cite{Hanai1960, Hanai1980, Hanai1986, Sen1981} to obtain
the complex permittivity of mixtures with different types of
inclusions (the so-called Bruggeman-Hanai or Maxwell-Wagner-Hanai
model). { It should be noted that the original Hanai's
generalization is based on the Maxwell-Wagner
model~\cite{Wagner1914, Hanai1960} and thus incorporates the
interfacial polarization effects (known as Maxwell-Wagner
polarization).} Generalizations of the differential approach to
bi-anisotropic systems were developed by Lakhtakia and
co-workers~\cite{Lakh1998, Michel2001} (incremental and
differential Maxwell-Garnett formalisms).

The term ``asymmetrical'' means that within the ABM, the system's
constituents are divided into ``the inclusions'' (filler
particles) and ``the matrix'' (host medium)\footnote{Obviously,
this division is not rigorous: as the volume concentration of the
particles becomes high, it is more natural to consider the host
medium as ``the filler'', and the filler particles as ``the
matrix''.}. To derive the differential equation for the effective
permittivity, the following procedure is envisaged. Given a
particular system, whose effective characteristics are formed by
the host and already added inclusions,  suppose that an
infinitesimal portion of inclusions is added to it. This addition
causes infinitesimal changes in the effective characteristics,
including the permittivity, of the system. Therefore,  another new
portion of inclusions will be added to a medium with a new
effective permittivity. So, as the desired system is built by
successive additions of infinitesimal portions of inclusions, each
new portion is added to a medium with its own effective
permittivity, different from that of the preceding medium.
Correspondingly, the portions added at different steps contribute
differently to the effective permittivity formation. The
differences between these contributions can be neglected for
diluted systems with low dielectric contrast, where the system's
constituents interact weakly. { For other
concentration ranges the ABM is, strictly speaking, inapplicable,
but we are unaware of any rigorous analysis of this matter.}

To get rid of the indicated limitations, each addition of a new
portion of inclusions should be analyzed by taking into account
the previously added inclusions~\cite{Chelidze}. This feature is
intrinsic to the symmetrical Bruggeman model
(SBM)~\cite{Bruggeman1936}, where all constituents of the system
are treated alike, and no differential equation is involved. Both
SBM and ABM methods belong to the class of effective-medium
approaches where one or all the constituents of the system are
embedded into some effective medium. However, the suggested ways
for modeling this medium are different in these methods. In the
SBM, the effective medium is formed by all the constituents
(including the real host). In differential models, the effective
medium is formed recursively, according to the Maxwell-Garnett
rule~\cite{M-G1904} {(or Maxwell-Wagner
rule~\cite{Wagner1914} if the permittivities are complex-valued)},
by successive addition of small portions of inclusions to the
current effective medium, starting from the pure matrix (for
details of the recursive procedure see, for
instance,~\cite{Lakh1998, Michel2001}). Nevertheless,
{ ignoring the above-indicated restriction on the
applicability ranges,} differential models are widely used for
electric spectroscopy studies of water/oil
emulsions~\cite{Hanai1960, Skodvin1994, Sjoblom1994, Jannier2013,
Pal2008}, soils~\cite{Shutko1982}, sands~\cite{Johnson2005} and
rocks~\cite{Sen1981, Chelidze1999}, and biological
samples~\cite{Bordi02, Yunus2002}, where the standard SBM does not
work properly.

{ The goal of this research is to scrutinize the
internal inconsistency and the ranges of validity of the classical
differential models for systems of particles with real- (the ABM
and its modifications) and complex-valued (the Maxwel-Wagner-Hanai
model and its modifications) permittivities. In order to do this
and avoid insignificant, within the scope of the research,
specific effects (such as absorption, interfacial Maxwell-Wagner
polarization), we develop a generalized differential approach to
the effective quasistatic permittivity of macroscopically
homogeneous and isotropic dielectric mixtures and then apply it to
the simplest system of impenetrable (``hard'') balls embedded in a
uniform host medium. The results obtained are, in fact, general
and applicable as long as the conditions of macroscopic
homogeneity and isotropy are fulfilled.}

The working model is based on the compact groups approach
(CGA)~\cite{Sushko2007, Sushko2009a, Sushko2009b, Sushko2013}. The
``compact groups'' are macroscopic regions in a system with
typical linear sizes $L$ that are much smaller than the wavelength
of probing field $\lambda$ in the system: $L \ll \lambda$. Such
groups are pointlike with respect to the field, but preserve the
effective properties of the system. This fact allows one to
effectively estimate the long-wavelength many-particle
contributions to the average electric field and induction in
macroscopically homogeneous and isotropic systems. In addition,
based on the CGA, some relations, essential for electrodynamic
homogenization and usually set \textit{ab initio}~\cite{Stroud75,
Mackay00, Mallet2005}, can be substantiated by using the boundary
conditions for the electric field (see \ref{app:theory})
or the Hashin-Shtrikman variational principle~\cite{HS1962} (see
\cite{Sushko2017}). {The analysis within the CGA actually
reduces to simple modeling of the dielectric permittivity
distribution in the system. }

{ Very recently~\cite{Sushko2017}, the CGA was used to
describe the effective dielectric response of dispersions of
graded impenetrable and hard-core--penetrable-shell particles. The
validity of the approach was demonstrated in \cite{Sushko2017} by
contrasting its results with existing rigorous analytic results
and computer simulations for dispersions of hard dielectric
spheres with power-law permittivity profiles, and by processing
experimental data on the effective dielectric response of
nonconducting polymer-ceramic composites. Earlier, the CGA was
efficiently applied to dispersions of particles with complex
permittivities to describe electric percolation phenomena in
composites of core-shell particles~\cite{Sushko2013}, two-step
electrical percolation in nematic liquid crystals filled with
multiwalled carbon nanotubes~\cite{Tomylko2015}, and effective
parameters of suspensions of nanosized insulating
particles~\cite{Sushko2016}. Finally, the idea of compact groups
was also used by Sushko to evaluate the effects of multiple
short-range reemissions between particles on the mean free path
and the transport mean free path of photons in concentrated
suspensions~\cite{SushkoJPS2009} and to discover the 1.5 molecular
light scattering in fluids near the critical
point~\cite{Sushko2004, SushkoCMP2013}; the results were supported
by extensive experimental data.}

{The above arguments show that the CGA is
well-substantiated, flexible, and efficient. For these reasons, it
was chosen as a basis for achieving the stated goals. The in-depth
analysis of the CGA for the model under consideration can be found
in \cite{Sushko2007, Sushko2009a,Sushko2009b,Sushko2017}.}

\section{General theoretical background}\label{sec:general}
The effective permittivity ${\varepsilon}$ is determined as the
proportionality coefficient between the average induction $\langle
\bf D \rangle$ and the average electric field $\langle \bf E
\rangle$:
\begin{equation}\label{eq:e_eff_def}
  \langle {\bf D}({\bf r}) \rangle
  = \epsilon_0 \langle \varepsilon ({\bf r}) {\bf E} ({\bf r}) \rangle
  = \epsilon_0 \varepsilon \langle {\bf E} ({\bf r}) \rangle,
\end{equation}
where $\epsilon_0$ is the electric constant, $\varepsilon({\bf r})
= \varepsilon_{\rm f} + \delta\varepsilon({\bf r})$ is the local
value of permittivity in the system, and the angular brackets
stand for statistical averaging. { For infinite
systems, the latter is equivalent to volume averaging,  according
to the ergodic hypothesis~\cite{LandauT8}.} The CGA allows one to
find these averages in the long-wavelength limit. For
macroscopically homogeneous and isotropic
systems~\cite{Sushko2007, Sushko2009a, Sushko2009b}
\begin{equation}\label{eq:E_average1-2}
  \left\langle {{\mathbf{E}}({\mathbf{r}})} \right\rangle
  = \left[ {1 + \sum\limits_{i = 1}^\infty  {{{\left( { - \frac{1}{{3{{ \varepsilon_{\rm f} }}}}} \right)}^i}} \left\langle {\delta {\varepsilon ^i}({\mathbf{r}})} \right\rangle } \right]{{\mathbf{E}}_0},
\end{equation}
\begin{equation}\label{eq:D_average1-2}
  \left\langle {{\mathbf{D}}({\mathbf{r}})} \right\rangle
  = \epsilon_0 { \varepsilon_{\rm f}} \left[ {1 - 2\sum\limits_{i = 1}^\infty  {{{\left( { - \frac{1}{{3{{ \varepsilon_{\rm f} }}}}} \right)}^i}} \left\langle {\delta {\varepsilon ^i}({\mathbf{r}})} \right\rangle } \right]{{\mathbf{E}}_0}.
\end{equation}
The functional form of $\delta\varepsilon({\bf r})$ is
constructed according to the system under
consideration. For a system of hard balls, with
permittivity $\varepsilon_1$, embedded in a host matrix, with
permittivity $\varepsilon_0$,  $\varepsilon({\bf r})$ is equal to
$\varepsilon_1$ in the regions occupied by the balls, and
$\varepsilon_0$ otherwise. Correspondingly,
$\delta\varepsilon({\bf r})$ takes the form
\begin{equation}\label{eq:delta-Brug}
  \delta\varepsilon_{\rm CGA} ({\bf r})
  = (\varepsilon_0 - \varepsilon_{\rm f}) \left( 1 - \tilde{\chi}_1 ({\bf r}) \right)
  + (\varepsilon_1 - \varepsilon_{\rm f}) \tilde{\chi}_1 ({\bf r}),
\end{equation}
where $\tilde{\chi}_1 ({\bf r})$ is the characteristic function of
the region occupied by all the balls. The parameter
$\varepsilon_{\rm f}$ is found in \ref{app:theory} and
in~\cite{Sushko2017}; it equals the effective permittivity:
$\varepsilon_{\rm f} = \varepsilon$.

In order to find $\varepsilon$, one should sum up the series in
(\ref{eq:E_average1-2}), (\ref{eq:D_average1-2}). For
(\ref{eq:delta-Brug}), this gives the following equation for
$\varepsilon$~\cite{Sushko2009a}:
\begin{equation}\label{eq:Bruggeman}
  (1 - c) \frac{\varepsilon_0 - \varepsilon}{2\varepsilon + \varepsilon_0} + c \frac{\varepsilon_1 - \varepsilon}{2\varepsilon + \varepsilon_1} = 0,
\end{equation}
where $c$ is a volume fraction of the inclusions phase.

According to the CGA, the long-wavelength results
(\ref{eq:E_average1-2}), (\ref{eq:D_average1-2}), and
$\varepsilon_{\rm f} = \varepsilon$ are valid for any
macroscopically homogeneous and isotropic system. Thus, they can
be applied to any distribution $\delta\varepsilon$
satisfying these conditions and are equivalent to the relation
(see \ref{app:theory} or~\cite{Sushko2017})
\begin{equation}\label{eq:general}
  \left\langle \frac{\delta\varepsilon({\bf r})}{3\varepsilon + \delta\varepsilon({\bf r})} \right\rangle = 0.
\end{equation}

The functional form of $\delta\varepsilon({\bf r})$ can vary
significantly. However, it can  always be categorized into two
classes, depending on the way in which the system's constituents
are treated: (1) symmetrical models (e.g. the SBM), where all the
constituents are treated equally; (2) asymmetrical models (e.g.
the ABM), where one phase is treated as ``the inclusion'' phase
and the other as ``the matrix'' phase. Some simplest forms of
$\delta\varepsilon$ and the corresponding mixing rules are as
follows:

1) The SBM where $\delta\varepsilon$ formally coincides with (\ref{eq:delta-Brug}) at $\varepsilon_{\rm f} = \varepsilon$.
This means that each inclusion and the matrix are embedded in an
effective medium with looked-for permittivity
$\varepsilon$. The homogenization condition
$\varepsilon_{\rm f} = \varepsilon$ is actually the
basic assumption of the model.  The SBM-type distributions
lead to the same result~(\ref{eq:Bruggeman}) for
$\varepsilon$. However, it should be noted that
the original SBM equation deals only with systems of hard
spheres. The constituents' polarizabilities
are identified with their individual polarizabilities in the
effective medium, and the matrix is assumed to polarize as
a single particle does~\cite{Banhegyi1986}. These two
suggestions, used also for systems of non-spherical particles,
are contradictory~\cite{Chelidze}.

In terms of the CGA, the form (\ref{eq:delta-Brug}) is the
most physically reasonable, since it represents the local
permittivity $\varepsilon({\bf r})$ in the system. As was
already mentioned, the latter is required to be
macroscopically homogeneous and isotropic, while the actual
form of the inclusions is insignificant. This fact
significantly distinguishes the CGA from the original SBM.

\begin{figure}[tb]
  \centering
  \includegraphics[height=0.17\textwidth]{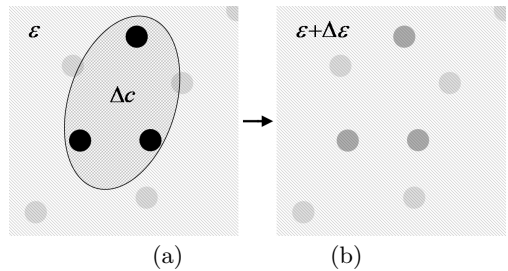}\\
  (a)\qquad\qquad\qquad(b)
  \caption{\label{fig:HanaiDiff}Schematic representation
  of the ABM differential algorithm: (a) addition
  of a portion of new particles with concentration
  $\Delta c/(1-c)$ in the particle-void region to the
  current effective medium with permittivity $\varepsilon$
  (the lighter area) leads to (b) the formation of a new
  effective medium with permittivity
  $\varepsilon + \Delta\varepsilon$,
  which serves as the matrix for the next portions of inclusions.
  Therefore, the previously added portions interact electrically
  with the new one only via the effective medium
  (comprising the particles shown darker).}
\end{figure}

2) The ABM can also be reproduced easily. Assume that the
effective permittivity $\varepsilon$ of the system at a
certain amount of inclusions is known (see
fig.~\ref{fig:HanaiDiff}(a)). If a small
portion of inclusions with the characteristic function
$\Delta {\tilde \chi}_1 ({\bf r})$ (${\tilde \chi}_1 \cdot
\Delta{\tilde \chi}_1 = 0$) is added (the enclosed area in
fig.~\ref{fig:HanaiDiff}(a)), the effective permittivity
will change by $\Delta \varepsilon$
(fig.~\ref{fig:HanaiDiff}(b)).
The current effective medium with permittivity
$\varepsilon$ is considered as the matrix
for the new inclusions that is void of any inclusions and
determined by the characteristic function
$(1 - {\tilde \chi}_1 - \Delta{\tilde \chi}_1)$.
In terms of $\delta\varepsilon$, this assumption can be written as
\begin{eqnarray}
  \delta\varepsilon_{\rm ABM}^{(l)} ({\bf r}) &=& (\varepsilon - (\varepsilon + \Delta\varepsilon)) [ 1 - {\tilde \chi}_1 ({\bf r}) - \Delta{\tilde \chi}_1 ({\bf r})] \nonumber\\
  &&+ (\varepsilon_1 - (\varepsilon + \Delta\varepsilon)) \Delta{\tilde \chi}_1({\bf r}) \nonumber\\
  &\approx& - \Delta\varepsilon [ 1 - {\tilde \chi}_1 ({\bf r}) ] + (\varepsilon_1 - \varepsilon) \Delta{\tilde \chi}_1({\bf r}),
\label{eq:delta-Hanai}
\end{eqnarray}
where only the terms of the first orders of smallness in $\Delta
{\tilde \chi}_1$ (in the meaning of its average) and
$\Delta\varepsilon$ are retained; $\varepsilon$ in
(\ref{eq:general}) also should be changed to $\varepsilon +
\Delta\varepsilon$, however it is not necessary for this
particular form of $\delta\varepsilon$. Substituting
(\ref{eq:delta-Hanai}) into (\ref{eq:general}), {
taking into account the ergodic hypothesis, and using the
orthogonality condition  $(1 - {\tilde \chi}_1 - \Delta{\tilde
\chi}_1)\Delta{\tilde \chi}_1 = 0$ for the characteristic
functions, averaging over the entire system in (\ref{eq:general})
can be split into averaging over the region occupied by the matrix
and averaging over the region occupied by a new portion of
inclusions:
\begin{eqnarray*}
  &&-\left\langle \frac{\Delta\varepsilon [1 - {\tilde \chi}_1 - \Delta {\tilde \chi}_1]}{3(\varepsilon + \Delta\varepsilon) + \Delta\varepsilon [1 - {\tilde \chi}_1 - \Delta {\tilde \chi}_1]} \right\rangle \nonumber\\
  &&\qquad + \left\langle \frac{(\varepsilon_1 - (\varepsilon + \Delta\varepsilon)) \Delta{\tilde \chi}_1}{3(\varepsilon + \Delta\varepsilon) + (\varepsilon_1 - (\varepsilon + \Delta\varepsilon)) \Delta{\tilde \chi}_1} \right\rangle \nonumber\\
  &&\approx - \frac{\Delta\varepsilon}{3\varepsilon} (1 - c) + \frac{\varepsilon_1 - \varepsilon}{2\varepsilon + \varepsilon_1}\Delta c = 0,
\end{eqnarray*}
where again only the terms of the first order of smallness were
retained. Changing to the infinitesimal values $d\varepsilon$ and
$dc$ gives the differential equation}
\begin{equation}\label{eq:Hanai-diff}
  \frac{d c}{1 - c} = \frac{d\varepsilon}{3\varepsilon} \frac{(2\varepsilon+\varepsilon_1)}{(\varepsilon_1 - \varepsilon)}.
\end{equation}
The point $c=1$ is a critical one; the solution of
(\ref{eq:Hanai-diff})  should satisfy there the condition
$\varepsilon = \varepsilon_1$. The ABM mixing rule
is obtained by integrating the left side of (\ref{eq:Hanai-diff})
with respect to $c$ from zero to $c$ and the right side with
respect to $\varepsilon$ from $\varepsilon_0$ to $\varepsilon$:
\begin{equation}\label{eq:Hanai}
  1 - c = \frac{\varepsilon - \varepsilon_1}{\varepsilon_0 - \varepsilon_1} \left( \frac{{ \varepsilon_0}}{\varepsilon} \right)^{1/3}.
\end{equation}
This equation has the same form as the one for the complex
permittivity (the integration is performed then  using Morera's and
Cauchy's theorems~\cite{Hanai1960}). As was mentioned in
Introduction, this approach is applicable for low
concentrations of inclusions (the upper index $l$ in
(\ref{eq:delta-Hanai}) is used to signify this fact).

In a similar  way, the high-concentration rule can be obtained.
The inclusions are now considered as ``the host medium'' and the
host medium as ``the inclusions'', with the characteristic function
$\tilde{\chi}_0 = (1 - \tilde{\chi}_1)$. A portion of ``the
inclusions'' with the characteristic function
$\Delta\tilde{\chi}_0 = - \Delta\tilde{\chi}_1$ is
embedded into that part of the current effective medium which
is void of ``the inclusions''; its characteristic function is
$(1 - \tilde{\chi}_0 - \Delta\tilde{\chi}_0)$. Correspondingly,
\begin{eqnarray}
  \delta\varepsilon_{\rm ABM}^{(h)} ({\bf r}) &\approx& - [ 1 - {\tilde \chi}_0 ({\bf r}) ] \Delta \varepsilon + (\varepsilon_0 - \varepsilon) \Delta{\tilde \chi}_0({\bf r}) \nonumber\\
  &=& - {\tilde \chi}_1 ({\bf r}) \Delta\varepsilon - (\varepsilon_0 - \varepsilon) \Delta{\tilde \chi}_1({\bf r}).
\label{eq:delta-Hanai-high}
\end{eqnarray}
Substituting (\ref{eq:delta-Hanai-high}) into (\ref{eq:general}) and
taking the necessary integrals give the desired rule:
\begin{equation}\label{eq:Hanai-high}
  c = \frac{\varepsilon - \varepsilon_0}{\varepsilon_1 - \varepsilon_0} \left( \frac{\varepsilon_1}{\varepsilon} \right)^{1/3}.
\end{equation}
Note that (\ref{eq:Hanai-high}) is used rarely, in contrast
to the original low-concentration ABM~\cite{Sen1981}.

3) The Looyenga~\cite{Looyenga1965} and
Lichtenecker~\cite{Licht1926} rules
\begin{equation}\label{eq:LO}
  \varepsilon^{1/3} = (1 - c) \varepsilon_0^{1/3} + c \varepsilon_1^{1/3},
\end{equation}
\begin{equation}\label{eq:LI}
  \log\varepsilon = (1 - c) \log\varepsilon_0 + c \log\varepsilon_1,
\end{equation}
for low-contrast systems~\cite{LandauT8, Simpkin2010}, can also
be obtained by substituting into (\ref{eq:general}) the formal expressions
\begin{eqnarray}
  \delta \varepsilon ({\bf r}) &=&
  (f(\varepsilon_0) - f(\varepsilon)) {(1 - {\tilde{\chi}_1 ({\bf r})})} \nonumber\\
  &&+ (f(\varepsilon_1) - f(\varepsilon)) {\tilde{\chi}_1 ({\bf r})},
\end{eqnarray}
with $f(x) = \{x^{1/3}, \log{x}\}$, respectively, and
keeping only the first-order terms in $|f(\varepsilon_i) - f(\varepsilon)|$ ($i=0,1$). However, these $\delta{ \varepsilon}$'s hardly
have transparent physical meanings.

\section{Differential scheme within the CGA}\label{sec:model}

In this section, we develop a general procedure for building
differential mixing rules based upon the  CGA. The above low-
(\ref{eq:Hanai}) and high-concentration (\ref{eq:Hanai-high})
rules turn out to be obtainable from the general differential
equation under certain simplifications. In other words, these ABM
rules are approximate  and of limited usefulness. The general
differential equation makes it possible not only to obtain their
improved modifications, but also investigate the validity limits
for the latter.

Suppose that an infinitesimal addition of inclusions to the system
causes the filler concentration and the effective permittivity to
change by small $\Delta c$ and $\Delta\varepsilon$,
respectively. In view of (\ref{eq:delta-Brug}), the new
permittivity distribution in the system becomes
\begin{eqnarray}
  \widetilde{\delta\varepsilon}_{\rm CGA} ({\bf r}) &=& (\varepsilon_0 - (\varepsilon + \Delta\varepsilon)) [1 - ({\tilde \chi}_1 ({\bf r}) + \Delta{\tilde \chi}_1 ({\bf r}))] \nonumber\\
  &&+ (\varepsilon_1 - (\varepsilon +   \Delta\varepsilon)) [{\tilde \chi}_1 ({\bf r}) + \Delta{\tilde \chi}_1 ({\bf r})],
\label{eq:delta-Brug-diff0}
\end{eqnarray}
and $\varepsilon$ in (\ref{eq:general}) changes to ${
\varepsilon} + \Delta\varepsilon$. Using simple algebraic
manipulations, the expression (\ref{eq:delta-Brug-diff0}) can be
represented as the sum of the distributions
(\ref{eq:delta-Brug}), (\ref{eq:delta-Hanai}), and
(\ref{eq:delta-Hanai-high}):
\begin{equation}\label{eq:delta-Brug-diff}
  \widetilde{\delta\varepsilon}_{\rm CGA} ({\bf r}) = \delta\varepsilon_{\rm ABM}^{(l)} ({\bf r}) + \delta\varepsilon_{\rm ABM}^{(h)} ({\bf r}) + \delta\varepsilon_{\rm CGA} ({\bf r}).
\end{equation}
Thus, according to the CGA, any changes in $\varepsilon$ caused by
the addition of small amounts of inclusions do not reduce to the
contribution from these inclusions alone (the term
$\delta\varepsilon_{\rm ABM}^{(l)}$ (\ref{eq:delta-Hanai}), as in
the ABM), but are also influenced by the changes in the host
volume fraction (the term $\delta\varepsilon_{\rm ABM}^{(h)}$
(\ref{eq:delta-Hanai-high})) and by the state of the system just
before the addition (the term $\delta\varepsilon_{\rm CGA}$
(\ref{eq:delta-Brug})). It follows immediately that the original
differential procedures \cite{Hanai1960, Hanai1980, Hanai1986,
Sen1981} behind the ABM are incomplete.

Substituting (\ref{eq:delta-Brug-diff}) into (\ref{eq:general})
and changing to the infinitesimal values, the differential
equation
\begin{eqnarray}
  &&\left[ d c \frac{\varepsilon_1 - \varepsilon}{2\varepsilon + \varepsilon_1} - (1 - c) \, d\varepsilon \frac{3\varepsilon_0}{(2\varepsilon + \varepsilon_0)^2} \right] \nonumber\\
  &&+ \left[ - d c \frac{\varepsilon_0 - \varepsilon}{2\varepsilon + \varepsilon_0}
  - c \, d\varepsilon \frac{3\varepsilon_1}{(2\varepsilon + \varepsilon_1)^2} \right] = 0.
\label{eq:Bruggeman-diff-general}
\end{eqnarray}
 is obtained. Actually, this is the differential form of (\ref{eq:Bruggeman}). It is convenient to use
(\ref{eq:Bruggeman-diff-general}) to derive new low- and
high-concentration modifications of the ABM rules.

Consider first the low concentration limit, where $c$ and $({
\varepsilon}_0 - \varepsilon)$ can be assumed to be of the same
order of smallness as $\Delta c$ and $\Delta\varepsilon$ are.
{Then $\delta\varepsilon_{\rm ABM}^{(h)}$ and the terms
in the second brackets in (\ref{eq:Bruggeman-diff-general}) are of
the second order of smallness and can be neglected.} Correspondingly,
 $\widetilde{\delta\varepsilon}_{\rm CGA}$ is determined only by
the first and third terms in (\ref{eq:delta-Brug-diff}):
\begin{equation}\label{eq:delta-Brug-diff-low-c}
    \widetilde{\delta\varepsilon}_{\rm CGA}^{(l)} \approx \delta\varepsilon_{\rm ABM}^{(l)} + \delta\varepsilon_{\rm CGA},
\end{equation}
and only the terms in the first brackets in
(\ref{eq:Bruggeman-diff-general}) must be retained.  This gives the
differential equation
\begin{equation}\label{eq:Bruggeman-diff-low-c}
    \frac{d c}{1-c} = d\varepsilon \frac{3\varepsilon_0(2\varepsilon + \varepsilon_1)}{(\varepsilon_1 - \varepsilon) (2\varepsilon + \varepsilon_0)^2}.
\end{equation}
{This equation can also be derived by directly
substituting (\ref{eq:delta-Brug-diff-low-c}) into
(\ref{eq:general}), retaining the terms of the first order of
smallness, and passing to the infinitesimal increments $dc$ and
$d\varepsilon$.}

An analogous procedure for the high-concentration limit gives
\begin{equation}\label{eq:delta-Brug-diff-high-c}
    \widetilde{\delta\varepsilon}_{\rm CGA}^{(h)} \approx \delta\varepsilon_{\rm ABM}^{(h)} + \delta\varepsilon_{\rm CGA},
\end{equation}
\begin{equation}\label{eq:Bruggeman-diff-high-c}
    \frac{d c}{c} = - d\varepsilon \frac{3\varepsilon_1 (2\varepsilon + \varepsilon_0)}{(\varepsilon_0 - \varepsilon) (2\varepsilon + \varepsilon_1)^2}.
\end{equation}

In general, equations (\ref{eq:delta-Brug-diff-low-c}) and
(\ref{eq:delta-Brug-diff-high-c}) differ considerably from their
ABM counterparts (\ref{eq:Hanai-diff}) and
(\ref{eq:delta-Hanai-high}), but reduce to them provided the term
$\delta\varepsilon_{\rm CGA}$ is neglected. This is possible
if: (1)  $\varepsilon_0 \approx \varepsilon$ and $\varepsilon_1
\approx \varepsilon$, respectively; (2) the concentration of the
constituent being added is small; (3)
$|\varepsilon_1-\varepsilon_0|$ is small as well.
It follows that the original ABM mixing rules are, in general,
physically inconsistent. In practice, one can attempt to apply
them only to diluted low-contrast systems.

\begin{figure}[tb]
    \centering
    \includegraphics[width=0.3\textwidth]{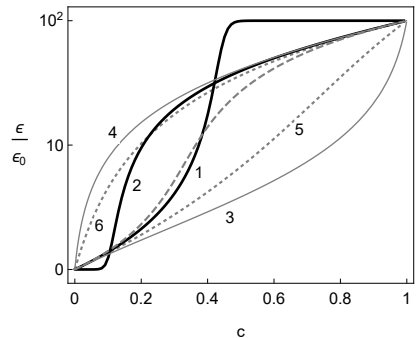}
    \caption{\label{fig:HSbounds}
    The concentration dependence of $\varepsilon$ according to: the new low-~(\ref{eq:Hanai-new})
    and high-concentration~(\ref{eq:Hanai-high-new}) rules (thick solid lines 1 and 2, respectively);
    Hashin-Shtrikman bottom (\ref{eq:HS-lower}) and upper (\ref{eq:HS-upper}) bounds (thin solid lines 3 and 4);
    CGA (\ref{eq:Bruggeman}) (dashed line); original  ABM  low-~(\ref{eq:Hanai}) and high-concentration~(\ref{eq:Hanai-high})
    rules (dotted lines 5 and 6). The only parameter used  $\varepsilon_1/\varepsilon_0 = 10^2$.}
\end{figure}

The equations (\ref{eq:Bruggeman-diff-low-c}) and
(\ref{eq:Bruggeman-diff-high-c}) are improved differential
equations,  which partially take into account, through
$\delta\varepsilon_{\rm CGA}$, the interaction between the
constituent.  The integration of them results in the following
mixing rules for low and high concentrations of the filler,
respectively:
\begin{eqnarray}
    \ln{(1-c)} &=& \frac{9 \varepsilon_0 \varepsilon_1}{(2\varepsilon_1 + \varepsilon_0)^2} \ln{\left[ \frac{3\varepsilon_0 (\varepsilon - \varepsilon_1)}{(\varepsilon_0 - \varepsilon_1) (2\varepsilon + \varepsilon_0)} \right]} \nonumber\\
    &&- \frac{2(\varepsilon_0 - \varepsilon_1) (\varepsilon_0 - \varepsilon)}{(2\varepsilon_1 + \varepsilon_0) (2\varepsilon + \varepsilon_0)};
\label{eq:Hanai-new}
\end{eqnarray}
\begin{eqnarray}
    \ln{c} &=& \frac{9 \varepsilon_0 \varepsilon_1}{(2\varepsilon_0 + \varepsilon_1)^2} \ln{\left[ \frac{3\varepsilon_1 (\varepsilon - \varepsilon_0)}{(\varepsilon_1 - \varepsilon_0) (2\varepsilon + \varepsilon_1)} \right]} \nonumber\\
    &&- \frac{2(\varepsilon_1 - \varepsilon_0) (\varepsilon_1 - \varepsilon)}{(2\varepsilon_0 + \varepsilon_1) (2\varepsilon + \varepsilon_1)}.
\label{eq:Hanai-high-new}
\end{eqnarray}
Compared to the ABM, we expect these rules to be more
accurate and valid for wider concentration regions. However, as
based on the assumptions (\ref{eq:delta-Brug-diff-low-c}) and
(\ref{eq:delta-Brug-diff-high-c}), they are still approximate. To
prove this fact, consider the Hashin-Shtrikman (HS) upper
$\varepsilon^{+}$ and bottom $\varepsilon^{-}$ bounds~\cite{HS1962}
\begin{equation}\label{eq:HS-upper}
    \varepsilon^{+} = \varepsilon_1 + \frac{3(1 - c)\varepsilon_1 (\varepsilon_0 - \varepsilon_1)}{3\varepsilon_1 + c
    (\varepsilon_0 - \varepsilon_1)},
\end{equation}
\begin{equation}\label{eq:HS-lower}
    \varepsilon^{-} = \varepsilon_0 + \frac{3c\varepsilon_0 (\varepsilon_1 - \varepsilon_0)}{3\varepsilon_0 + (1-c) (\varepsilon_1 - \varepsilon_0)}.
\end{equation}
It is easy to see that the rules (\ref{eq:Hanai-new}) and
(\ref{eq:Hanai-high-new}) fail to satisfy these bounds (see
Fig.~\ref{fig:HSbounds}). Indeed, consider the rule
(\ref{eq:Hanai-new}) for a system with $\varepsilon_1 \gg
\varepsilon_0$. For  $c > (1 - e^{-1/2}) \approx 0.393$,
$\varepsilon \to \varepsilon_1$, which is higher than the upper HS
bound (\ref{eq:HS-upper}) for the same concentrations. In the
region of low concentrations (\ref{eq:Hanai-new}) is closer to
(\ref{eq:Bruggeman}) than (\ref{eq:Hanai}) and falls in between
the  HS bounds. Similarly, for (\ref{eq:Hanai-high-new}) and $c <
e^{-2} \approx 0.135$, $\varepsilon \to \varepsilon_0$, which is
lower than the HS bottom bound (\ref{eq:HS-lower}).

For arbitrary $\varepsilon_1$ and $ \varepsilon_0$ the
concentrations where the HS bounds are violated depend on the
contrast $\varepsilon_0/\varepsilon_1$. Figure ~\ref{fig:HSbounds}
illustrates the situation for a system with
$\varepsilon_1/\varepsilon_0 = 10^2$. It is interesting to note
that the original ABM rules (\ref{eq:Hanai}) and
(\ref{eq:Hanai-high}) satisfy the HS bounds. According to the
above discussion, this fact is not indicative of the superiority of the ABM
rules (\ref{eq:Hanai}) and (\ref{eq:Hanai-high}) over their
modifications (\ref{eq:Hanai-new}) and (\ref{eq:Hanai-high-new}),
but reflects the changes in the interplay between $ \delta{
\varepsilon}_{\rm ABM}^{(l)} ({\bf r})$, $ \delta{
\varepsilon}_{\rm ABM}^{(h)} ({\bf r})$, and $ \delta{
\varepsilon}_{\rm CGA} ({\bf r}) $ in the formation of
$\varepsilon$ that occur as $c$ is changed. In other words, a
simple extrapolation of the simplifications used within the
differential method for one narrow concentration range fails to
incorporate the effects essential for the formation of
$\varepsilon$ in other concentration ranges.

{ It should be noted that the above results
quantitatively support the well-known qualitative
arguments~\cite{Chelidze, Chelidze1999} that at high
concentrations, both the ABM and Maxwell-Wagner-Hanai models do
not fully take into account interparticle polarization effects.
They also explain why it is necessary to modify the classical
differential models, or even introduce additional adjustable
parameters, in order to extend their ranges of
applicability~\cite{Becher1987, Sihvola2007}. And they are in accord the
final-element calculations \cite{Mejdoubi2007} which show that at
small concentrations, the small changes of the effective
permittivity due to the addition of a new portion of inclusions
are greater than those predicted by the differential mixing
procedure. }

\section{Conclusion}\label{sec:conclusion}
The differential approach to the effective permittivity
$\varepsilon$ of a system implies that infinitesimal changes in
the concentrations of constituents lead to infinitesimal changes
in $\varepsilon$; the result is a differential equation for
$\varepsilon$. In the present work, we develop a general
differential scheme for dielectric systems of impenetrable balls
based upon the compact groups approach (CGA). For this purpose,
the CGA is formulated in a way that allows one to analyze the role
of different contributions to the model permittivity distribution
in a system. The analysis of these contributions
 and corresponding differential equations
reveals that:

\begin{enumerate}
\item{The low- and high-concentration mixing rules of the
classical asymmetrical Bruggeman model (ABM) are reproducible
within our model under the suggestion that the electromagnetic
interaction between the inclusions already contained in the system
with those being added can be replaced by the interaction
of the latter with the current effective medium. Therefore, the
classical ABM mixing rules are, in general, physically
inconsistent and applicable only for diluted (with respect to
one of the constituents) systems with low dielectric contrast
between the constituents.}

\item {The overall changes in $\varepsilon$ due to addition of an
infinitesimal portion of one constituent include the contributions
from both constituents (inclusions and the host medium) and depend
on the state of the system before the addition. Ignoring the
contribution from one of the constituents, we obtain generalized
versions of the original ABM mixing rules.}

\item {The new generalized differential mixing rules are, again,
applicable only in certain concentration ranges because beyond
those they do not satisfy the Hashin-Shtrikman
bounds. This means that  different mechanisms are responsible for
the formation of $\varepsilon$ in different concentration ranges.
Simple extrapolation of the results obtained for a certain
concentration range cannot incorporate all the effects essential
for the formation of $\varepsilon$ in the whole concentration
range.}
\end{enumerate}

The results obtained can be generalized to macroscopically
homogeneous and isotropic systems with complex permittivities of
the constituents\footnote{It should be noted that, when applied to
a system of particles with complex permitivities, the CGA is
capable of giving quantitative estimates of such specific effects
as interfacial polarization. Indeed, the CGA generalizes the ABM
which, being a generalization (known as the Maxwell-Wagner-Hanai
model) of the Maxwell-Wagner model (see Introduction) to
complex-valued permittivities, automatically incorporates these
effects. We plan to present these results elsewhere.}.
{ However, it should be emphasized that the novelty of
this article is not a derivation of new differential mixing rules,
but the quantitative proof of a general statement that
differential mixing rules are approximate and applicable only for
narrow ranges of concentration and dielectric contrast. Attempts
to go beyond those ranges will be, strictly speaking, misleading
because of uncontrollable mistakes inherent to the differential
method.}

\section*{Acknowledgement}
I thank Dr.~M.~Ya.~Sushko for helpful discussions and
encouragement and Prof.~A.~Lakhtakia for drawing my attention
to incremental and differential Maxwell-Garnett formalisms.

\section*{Appendix}
\appendix

\section{Derivation of equation (\ref{eq:general})}\label{app:theory}
The first step is to present the equation
\begin{equation}\label{eq:wave_eq}
      \Delta \mathbf{E}(\mathbf{r}) - \nabla(\nabla \mathbf{E}(\mathbf{r})) + k_0^2 \epsilon_0 \varepsilon_{\rm f} \mathbf{E}(\mathbf{r}) = - k_0^2 \epsilon_0 \delta\varepsilon(\mathbf{r}) \mathbf{E}(\mathbf{r})
\end{equation}
for a wave propagating in the inhomogeneous  medium, with local
permittivity $\varepsilon({\bf r}) = \varepsilon_{\rm f} +
\delta\varepsilon({\bf r})$, in the integral form
\begin{equation}\label{eq:wave_prop_int}
\mathbf{E}(\mathbf{r}) = \mathbf{E}_0(\mathbf{r}) - k_0^2
\epsilon_0 \int\limits_V {d{\mathbf{r}}'}
     {\rm{\hat T}}({\mathbf{r}} - {\mathbf{r}}')\delta \varepsilon ({\mathbf{r}}'){\mathbf{E}}({\mathbf{r}}').
\end{equation}
Here: $\nabla$ is the del operator; $\Delta$ is the Laplace
operator; the integral is taken over the volume $V$ of the system
under consideration;  ${\bf E}_0 ({\bf r}) = {\bf E}_0 \exp{ ({\rm
i} \sqrt{\varepsilon_{\rm f}} {\bf k}_0 \cdot {\bf r}) }$ is a
probing field with amplitude $\bf{E}_0$ and wave vector
${\bf k} = \sqrt{\varepsilon_{\rm f}} {\bf k}_0 $ in the medium;
$\varepsilon_{\rm f}$ is the permittivity of an unknown
background medium, in which each constituent
(including the host medium) is embedded; $\rm{\hat T}$ is the
electromagnetic field propagator (the Green's tensor for
(\ref{eq:wave_eq})). It can be shown \cite{Sushko2004, Sushko2007,
Weiglhofer1995} that  $\rm{\hat T}$ can be associated with the
tensor $\rm {\widetilde T}$, such that
\[
  \int\limits_V {d{\bf r}\, {\rm{\hat T}} ({\bf r}) \psi({\bf r})} = \int\limits_V {d{\bf r}\, {\rm{\widetilde T}} ({\bf r}) \psi({\bf r})}
\]
for ``sufficiently good'' bounded scalar
functions $\psi({\bf r})$. In the long-wave limit ($|{\bf k}| \to 0$),
the components of $\rm {\widetilde T}$ satisfy the
relation
\begin{eqnarray}
  k_0^2 \epsilon_0 {\widetilde T}_{\alpha \beta} ( {\mathbf{r}} ) = {\widetilde T}_{\alpha \beta}^{(1)} ( {\mathbf{r}} ) &+& {\widetilde T}_{\alpha \beta}^{(2)}( {\mathbf{r}} ) = \frac{1}{3 \varepsilon_{\rm f}}\delta ({\mathbf{r}}) \delta_{\alpha \beta} \nonumber\\
  &+& \frac{1}{4\pi \varepsilon_{\rm f} r^3} \left( \delta_{\alpha \beta} - 3 \frac{r_{\alpha} r_{\beta}}{r^2} \right),
\label{eq:prop_longwave}
\end{eqnarray}
where $\delta(\mathbf{r})$ is the Dirac delta-function, and
$\delta_{\alpha \beta}$ is the Kronecker delta. The part
$\rm{\widetilde T}^{(1)}$ characterizes the
effects of multiple reemissions within compact
groups~\cite{Sushko2007}.

Substituting (\ref{eq:prop_longwave}) into
(\ref{eq:wave_prop_int}), making elementary algebraic
manipulations, and averaging statistically, we obtain the relation
\begin{eqnarray}
    \langle \mathbf{E} && (\mathbf{r}) \rangle = \left\langle \frac{ 3 \varepsilon_{\rm f}  }{3 \varepsilon_{\rm f} + \delta\varepsilon({\bf r})} \right\rangle \mathbf{E}_0 \nonumber\\
    &&- 3 \varepsilon_{\rm f} \int\limits_V {d{\mathbf{r}}'} { \rm{\widetilde T}^{(2)} (| {\bf r} - {\bf r}' |)     \left\langle \frac{\delta \varepsilon ({\mathbf{r}}')}{3 \varepsilon_{\rm f} + \delta \varepsilon({\bf r})}{\mathbf{E}}({\mathbf{r}}') \right\rangle } .
\label{eq:wave_prop_int2}
\end{eqnarray}
The local deviation $\delta\varepsilon$ here corresponds to
macroscopic compact groups~\cite{Sushko2007, Sushko2009b}, not
single particles (as in microscopic approaches, such as
\cite{Stroud75, Mackay00, Mallet2005}). For macroscopically
isotropic and homogeneous systems, the
statistical average in the integrand depends only on $|{\bf r} -
{\bf r}'|$. Because of this symmetry and  a specific angular
dependence of $\widetilde{T}_{\alpha\beta}^{(2)}$, the second term
in the right-hand side of (\ref{eq:wave_prop_int2}) vanishes, and
(\ref{eq:wave_prop_int2}) reduces to
\begin{equation}\label{eq:E_average2}
  \langle {\bf E} ({\bf r}) \rangle =  \xi {\bf E}_0,\quad
  \xi  = \left\langle { \frac{3 \varepsilon_{\rm f}}{3 \varepsilon_{\rm f} + \delta\varepsilon({\bf r})} } \right\rangle.
\end{equation}

The average displacement $\langle{\bf D} ({\bf r})\rangle$ is
found using definition (\ref{eq:e_eff_def}) and applying the
above symmetry reasoning to the
$\widetilde{T}_{\alpha\beta}^{(2)}$-involving integral. The result
is:
\begin{equation}\label{eq:D_average2}
  \langle {\bf D} ({\bf r}) \rangle = \epsilon_0 \varepsilon_{\rm f} \xi {\bf E}_0 + \epsilon_0 \langle \delta\varepsilon {\bf E}({\bf r}) \rangle
  = \epsilon_0 \varepsilon_{\rm f} (1 + 2\eta) {\bf E}_0,
\end{equation}
\[
  \eta = \left\langle { \frac{\delta\varepsilon({\bf r})}{3 \varepsilon_{\rm f} + \delta \varepsilon({\bf r})} } \right\rangle,
\]
where it was taken into account that
\begin{equation}\label{eq:xi-eta_rel}
  \xi + \eta = 1.
\end{equation}

Note that by representing (\ref{eq:E_average2}) and
(\ref{eq:D_average2}) as infinite geometric sequences, we recover
the iteration series (\ref{eq:E_average1-2}) and
(\ref{eq:D_average1-2})~\cite{Sushko2007, Sushko2009b}.

Using (\ref{eq:e_eff_def}) and taking into account
(\ref{eq:E_average2}), (\ref{eq:D_average2}), and
(\ref{eq:xi-eta_rel}), we obtain
\begin{equation}\label{eq:eta}
  \varepsilon - \varepsilon_{\rm f}=(\varepsilon +2 \varepsilon_{\rm
  f})\eta.
\end{equation}

In order to find the unknown $\varepsilon_{\rm f}$, we need one
more relation between $\varepsilon$ and $\varepsilon_{\rm
f}$.  We derive it using the equations between the electric field
in vacuum and those in the effective and background media:
\[
  {\bf E}_{\rm vac} = \varepsilon \left\langle \bf E \right\rangle, \quad{\bf E}_{\rm vac} = \varepsilon_{\rm f} {\bf E}_{0}.
\]
Then
\[
  \varepsilon_{\rm f} {\bf E}_{0} = \varepsilon \left\langle {\bf E} \right\rangle = \varepsilon \xi {\bf E}_{0},
\]
whence, in view of (\ref{eq:xi-eta_rel}),
\[
  \varepsilon - \varepsilon_{\rm f} = \varepsilon\eta.
\]
This and (\ref{eq:eta}) give the relation
\begin{equation}\label{eq:eps_f_rel_homog}
  \eta = \frac{\varepsilon - \varepsilon_{\rm f}}{2\varepsilon_{\rm f} + \varepsilon} = \frac{\varepsilon - \varepsilon_{\rm f}}{\varepsilon}.
\end{equation}
Of its two roots $\varepsilon_{\rm f} = 0$ and ${
\varepsilon}_{\rm f} = \varepsilon$, only the latter is
physically meaningful. It corresponds to the well-known
Bruggeman-type homogenization and gives the equality
$\eta|_{\varepsilon_{\rm f} = \varepsilon} = 0$, that is,
equation (\ref{eq:general}).

It should be emphasized that for  macroscopically homogeneous and
isotropic systems, considered in the long-wavelength limit,
the homogenization condition $\varepsilon_{\rm f} = {
\varepsilon}$ is independent of the functional form of $\delta{
\varepsilon}$. The same result was obtained in~\cite{Sushko2017}
by combining the CGA with the Hashin-Shtrikman variational theorem
\cite{HS1962}. This condition is the key relation postulated in
many theories, including the strong-property-fluctuation
theory~\cite{Mackay00}, where it is used to get rid of the secular
terms in the Born series for the renormalized electric field.

%\bibliographystyle{unsrt}
%\bibliography{refs}

\end{document}